\documentclass[aps,prb,twocolumn,showpacs,preprintnumbers,amsmath,amssymb,superscriptaddress]{revtex4}%

\usepackage{graphicx}
\usepackage{dcolumn}
\usepackage{bm}
\usepackage{color}

\begin{document}

\preprint{preprint(\today)}

\title{Strong coupling between the Eu$^{2+}$ spins and the Fe$_{2}$As$_{2}$ layers in EuFe$_{1.9}$Co$_{0.1}$As$_{2}$ observed with NMR}

\author{Z.~Guguchia}
\email{zurabgug@physik.uzh.ch} \affiliation{Physik-Institut der
Universit\"{a}t Z\"{u}rich, Winterthurerstrasse 190, CH-8057
Z\"{u}rich, Switzerland}

\author{J.~Roos}
\affiliation{Physik-Institut der Universit\"{a}t Z\"{u}rich,
Winterthurerstrasse 190, CH-8057 Z\"{u}rich, Switzerland}

\author{A.~Shengelaya}
\affiliation{Department of Physics, Tbilisi State University,
Chavchavadze 3, GE-0128 Tbilisi, Georgia}

\author{S.~Katrych}
\affiliation{Laboratory for Solid State Physics, ETH Z\"urich,
CH-8093 Z\"{u}rich, Switzerland}

\author{Z.~Bukowski}
\affiliation{Laboratory for Solid State Physics, ETH Z\"urich,
CH-8093 Z\"{u}rich, Switzerland}

\author{S.~Weyeneth}
\affiliation{Physik-Institut der Universit\"{a}t Z\"{u}rich,
Winterthurerstrasse 190, CH-8057 Z\"{u}rich, Switzerland}

\author{F.~Mur\'{a}nyi}
\affiliation{Physik-Institut der Universit\"{a}t Z\"{u}rich,
Winterthurerstrasse 190, CH-8057 Z\"{u}rich, Switzerland}

\author{S.~Str\"{a}ssle}
\affiliation{Physik-Institut der Universit\"{a}t Z\"{u}rich,
Winterthurerstrasse 190, CH-8057 Z\"{u}rich, Switzerland}

\author{A.~Maisuradze}
\affiliation{Physik-Institut der Universit\"{a}t Z\"{u}rich,
Winterthurerstrasse 190, CH-8057 Z\"{u}rich, Switzerland}
\affiliation{Laboratory for Muon Spin Spectroscopy, Paul Scherrer Institute, CH-5232
Villigen PSI, Switzerland}

\author{J.~Karpinski}
\affiliation{Laboratory for Solid State Physics, ETH Z\"urich,
CH-8093 Z\"{u}rich, Switzerland}

\author{H.~Keller}
\affiliation{Physik-Institut der Universit\"{a}t Z\"{u}rich,
Winterthurerstrasse 190, CH-8057 Z\"{u}rich, Switzerland}

\begin{abstract}
A combination of X-ray diffraction, magnetization, and
$^{75}$As nuclear magnetic resonance (NMR) experiments were performed on single-crystal
EuFe$_{1.9}$Co$_{0.1}$As$_{2}$. The strength of the hyperfine
interaction between the $^{75}$As nuclei and the Eu$^{2+}$ 4\textit{f} states
suggests a strong coupling between the Eu$^{2+}$ moments and the
Fe$_{1.9}$Co$_{0.1}$As$_{2}$ layers. Such a strong interlayer coupling
may be due to an indirect exchange interaction between the
localized Eu$^{2+}$ 4\textit{f} moments, mediated by the Fe 3\textit{d} conduction electrons.
Magnetic susceptibility as well as 
$^{75}$As-NMR measurements reveal a decrease of the
SDW transition temperature to $T_{\rm SDW}$=120 K as a result of Co-doping.
A change
of the slope in the temperature dependence of the NMR frequency of the $^{75}$As
lower-satellite line was observed at 225 K. At the same temperature
also a change of the satellite
line shape was found. These changes of the NMR spectra may be caused
by the formation of a nematic phase below 225 K in EuFe$_{1.9}$Co$_{0.1}$As$_{2}$.

\end{abstract}

\pacs{74.70.-b, 76.60.-k, 75.30.Fv, 74.25.Jb}

\maketitle

\section{Introduction}
The discovery  of superconductivity in iron-based
arsenides at temperatures  up to 56 K \cite{Kamihara08,Chen08,ChenWu08,Ren08,RenYang08,Wen08,Wang08} has  triggered  extensive interest in their
physical properties and the underlying mechanism of high-temperature superconductivity. The undoped parent compounds adopt a tetragonal
structure at room temperature, which consists of
[Fe$_{2}$As$_{2}$]$^{2-}$ layers separated by
[\textit{Ln}$_{2}$O$_{2}$]$^{2+}$ (\textit{Ln}=lanthanoide) layers \cite{Johnson74,Quebe2000} or \textit{A}$^{2+}$ (\textit{A}=Ca, Sr, Ba, Eu)
layers. \cite{Pfisterer80,Pfisterer83,Marchand78,Wu08} At low temperatures, the parent compounds undergo a
structural phase transition from a tetragonal to an orthorhombic phase,
accompanied \cite{Rotter08} or followed \cite{deLaCruz08} by a spin density wave (SDW) transition
 of the itinerant Fe moments. The superconducting (SC) state  can
be achieved  either  by  electron or by hole  doping  of  the  parent
compounds, leading to a suppression of the SDW formation. \cite{Wen08,RenLu08,Matsuishi08,Zhao08} The suppression of the magnetic transition in connection with the
simultaneous formation of a SC state is reminiscent  of cuprates and heavy  fermion  systems, therefore
suggesting that the SC state  in  these  systems  is unconventional as well.

 EuFe$_{2}$As$_{2}$ is a particularly interesting member of the iron
arsenide \textit{A}Fe$_{2}$As$_{2}$ ('122') family, since the \textit{A} site is occupied by a
Eu$^{2+}$ \textit{S}-state (orbital moment \textit{L}=0) rare-earth ion with a 4\textit{f}$^{7}$ electronic 
configuration with a total electron spin \textit{S}=7/2, corresponding to a theoretical effective
magnetic moment of 7.94 $\mu$$_{B}$. Figure~1 shows the crystal structure of EuFe$_{2}$As$_{2}$.
This compound is built up by [FeAs]$^{2-}$ layers, seperated by layers of magnetic Eu$^{2+}$ ions.
EuFe$_{2}$As$_{2}$ exhibits both a SDW ordering of the Fe moments and an antiferromagnetic ordering  of the localized
Eu$^{2+}$ moments below 190 K and 19 K, respectively. The presence of
magnetic phase transitions at 19 K and 190 K in EuFe$_{2}$As$_{2}$
was seen by M\"{o}ssbauer spectroscopy \cite{Raffius93} and is confirmed 
by neutron diffraction. \cite{Xiao09}  In contrast to the
other '122' systems, where the substitution of Fe by Co leads to
superconductivity, \cite{Sefat,Jasper} the compounds containing Eu$^{2+}$ exhibit the onset of a superconducting
transition but seem to be hindered to reach zero resistivity at ambient pressure. \cite{He08}
Reentrant superconducting behavior was also
observed in a EuFe$_{2}$As$_{2}$ crystal under applied pressure up to 2.5 GPa. \cite{Miclea09,Terashima} Only above 2.8 GPa a sharp resistive transition to a zero-resistivity state is achieved. \cite{Terashima} Bulk superconductivity is also observed in EuFe$_{2}$As$_{2-x}$P$_{x}$, \cite{Sun,Jeevan} where isovalent P-substitution of the As-site acts as chemical pressure on EuFe$_{2}$As$_{2}$.
No superconductivity was detected in EuFe$_{2-x}$Ni$_{x}$As$_{2}$, \cite{ZRen09} while
superconductivity with a maximum $T_{\rm c}$ ${\simeq}$ 20 K was
reported for BaFe$_{2-x}$Ni$_{x}$As$_{2}$. \cite{LiLuo08} It was
suggested from different experiments \cite{Jiang09,Dengler10,Ying10,ZRen09} that there is a strong coupling between the localized Eu$^{2+}$
spins and the conduction electrons of the two-dimensional (2D) Fe$_{2}$As$_{2}$
layers.
The study of the interaction between  the rare-earth Eu$^{2+}$ magnetic moments and
the conducting Fe$_{2}$As$_{2}$ layers is important in order to understand
why it is difficult to induce superconductivity in Co-doped EuFe$_{2-x}$Co$_{x}$As$_{2}$. 
\begin{figure}[t!]
\includegraphics[width=1.1\linewidth]{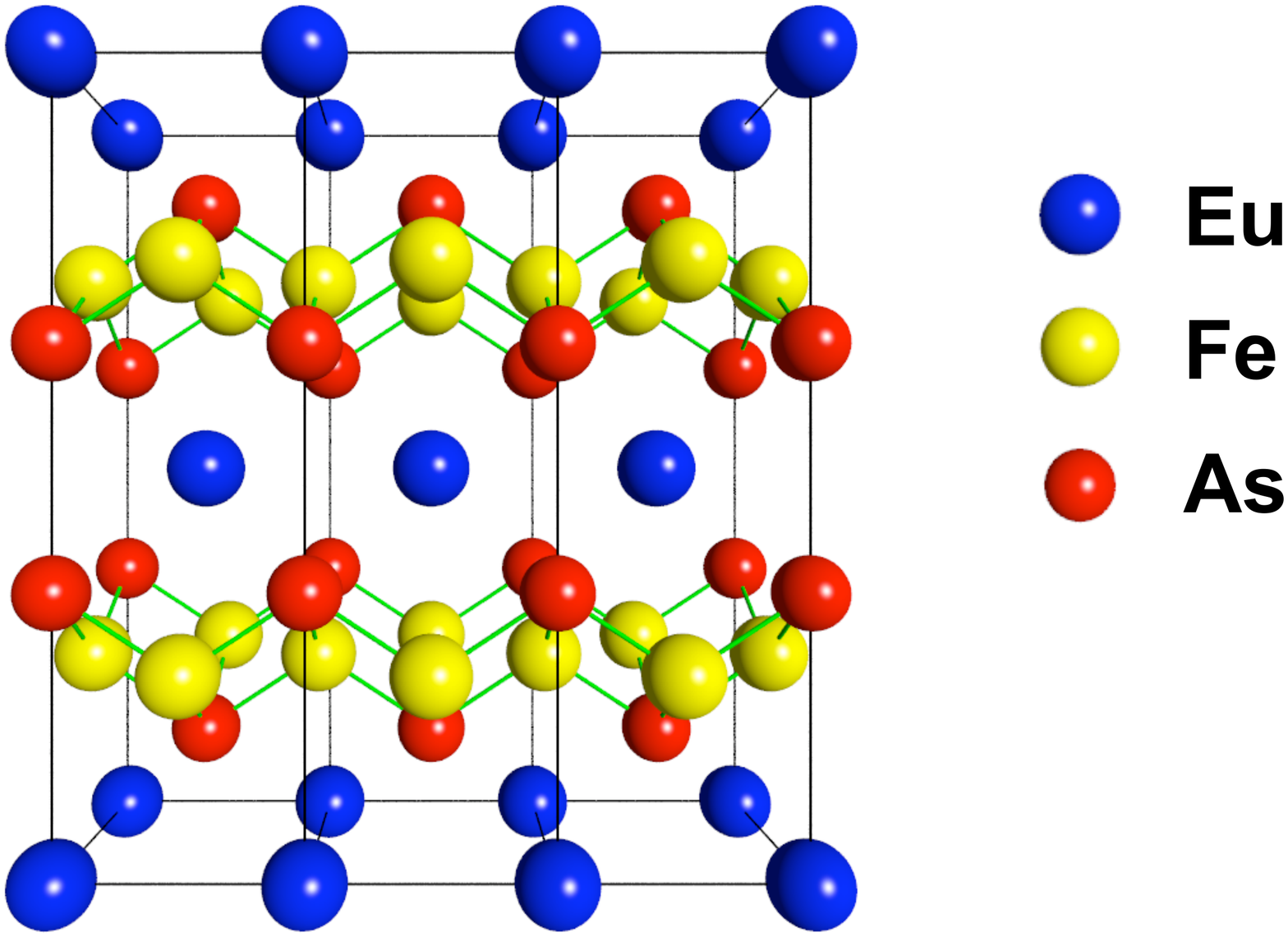}
\caption{ (Color online) Tetragonal crystal structure of EuFe$_{2}$As$_{2}$ at room temperature, consisting of
[Fe$_{2}$As$_{2}$]$^{2-}$ layers separated by Eu$^{2+}$ layers.}
\label{fig6}
\end{figure}
  In order to investigate the coupling between the Eu and
Fe$_{1.9}$Co$_{0.1}$As$_{2}$ layers as well as to study 
the magnetic transitions in EuFe$_{1.9}$Co$_{0.1}$As$_{2}$, a combination of X-ray diffraction, magnetization, and $^{75}$As nuclear magnetic resonance (NMR) experiments
were performed on single crystals. Magnetic susceptibility as well as 
$^{75}$As-NMR measurements reveal a decrease of the
SDW transition temperature to $T_{\rm SDW}$=120 K for EuFe$_{1.9}$Co$_{0.1}$As$_{2}$.
It was found that the $^{75}$As NMR spectra
are characterized by a negative frequency shift with respect to the
$^{75}$As-NMR Larmor frequency. This shift is significantly
larger than the one observed in \textit{A}Fe$_{2}$As$_{2}$ (\textit{A}=Ba, Ca, Sr).
\cite{Kitagawa1,Kitagawa2,Ning08,Baek09} The temperature dependence of the shift follows a
Curie-Weiss type behavior with a Curie-Weiss temperature close to
the one determined from the magnetization data. The estimate of the hyperfine
coupling constant between the $^{75}$As nuclei and the Eu 4\textit{f} states suggests a strong coupling between the Eu$^{2+}$ magnetic moments and the Fe$_{1.9}$Co$_{0.1}$As$_{2}$ layers. 



\section{EXPERIMENTAL DETAILS }

Single crystals of EuFe$_{1.9}$Co$_{0.1}$As$_{2}$ were grown out
of Sn flux. The chemical composition of the single crystals was 
determined on freshly cleaved samples using wavelength-dispersive
X-ray spectroscopy (WDS). The obtained composition 
corresponds to the formula EuFe$_{1.9}$Co$_{0.1}$As$_{2}$ within
the experimental error (${\pm}$ 5 $\%$). 
X-ray diffraction was performed on a single
crystal of a size of approximately 0.2 x 0.1 x 0.008 mm$^3$ using a
Bruker diffractometer equipped with the APEX II CCD detector (Bruker
ASX). The data were analyzed using the APEX2 \cite{APEX2}
and SAINT \cite{SAINT} software. The susceptibility measurements of
the crystals EuFe$_{1.9}$Co$_{0.1}$As$_{2}$ were carried out with a
SQUID magnetometer ($Quantum Design$) in the temperature range from
5 to 300 K in a magnetic field of 0.3 T applied parallel (\textit{H} ${\parallel}$ \textit{c}) and
perpendicular (\textit{H} ${\perp}$ \textit{c}) to the crystallographic \textit{c}-axis. The $^{75}$As-NMR
experiments on a single crystal (dimensions: 4 x 4 x 0.2 mm$^3$) from
the same batch were performed in an external magnetic field of 9 T
using a standard pulse spectrometer. NMR-echo signals were recorded
with a frequency selective echo pulse sequence  applying a phase-alternating
add-subtract accumulation technique. The $^{75}$As-NMR spectra
were obtained by scanning the frequency in discrete steps and
integrating the spin-echo signal, yielding the 'spin-echo
intensity'. The spin-spin relaxation time $T_{\rm 2}$ was
determined by measuring the spin-echo intensity as a function of the
delay time between the exciting and the refocusing pulse. 

\section{RESULTS AND DISCUSSION}
\subsection{Single crystal X-ray diffraction}
 X-ray diffraction experiments at room temperature
revealed a good quality of the EuFe$_{1.9}$Co$_{0.1}$As$_{2}$
crystal.
The average mosaic spread was estimated to be ${\simeq}$ 0.9$^{\circ}$. The
lattice constants for the tetragonal unit cell based upon the refinement of
689 reflections
are ${\textit a}$ = ${\textit b}$ = 3.9104(1) \AA, ${\textit c}$ =
11.9434(3) \AA, ${\textit V}$ = 182.629(8) \AA$^{3}$. The
average residual for symmetry equivalent reflections is $R_{\rm
int}$ = 4.63  ${\%}$  and  $R_{\rm \sigma}$ = 4.01 ${\%}$. The
\begin{table}[b!]
\caption{Crystallographic and structure refinement parameters of
the single crystal EuFe$_{1.9}$Co$_{0.1}$As$_{2}$. 
The diffraction study was performed at 295(2) K using MoK$_{\alpha}$
radiation with ${\lambda}$ = 0.71073 \AA. The lattice is
tetragonal, ${\textit I}$4/${\textit mmm}$ space group with ${\textit Z}$ = 2, atomic
coordinates: Eu on 2${\textit a}$ (0, 0, 0), Fe/Co on 4${\textit
d}$ (0, 1/2, 1/4), As on 4${\textit e}$ (0, 0, $z_{\rm As}$). A
full-matrix least-squares method was employed to optimize $F^{2}$.}
\vspace{0.3cm}
\begin{tabular}{l*{20}{l}r}
\hline
Empirical formula                          & EuFe$_{1.9}$Co$_{0.1}$As$_{2}$ \\
Unit cell dimensions (\AA)                 & \textit{a} = 3.9104(1)\\ 
                                           & \textit{c} = 11.9434(3)\\                             
Volume (\AA$^{3}$)                         & 182.629(8)     \\
$z_{\rm As}$ (atomic coordinate)           & 0.6388(1)       \\
$h_{pn}$ (\AA)                             & 1.3286                     \\
Calculated density (g/cm$^{3}$)            & 7.519                          \\
Absorption coefficient (mm$^{-1}$)         & 42.512                         \\
Absorption correction type                 & Numerical                      \\  
                                           & (from face indices)       \\
\textit{F}(000)                            & 362                            \\
Crystal size (${\mu}$m$^{3}$)              & 200 x 100 x 8                  \\
Theta range for data collection (deg)      & 3.41 to 43.92                  \\
Index ranges                               & -7 ${\leq}$ \textit{h} ${\leq}$ 5     \\
                                           & -5${\leq}$ \textit k ${\leq}$ 7  \\ 
                                           & -22${\leq}$ \textit l ${\leq}$ 17 \\        
Reflections collected/unique               & 884/243 $R_{\rm int}$ = 0.0463    \\ 
Completeness to 2${\Theta}$                & 97.2                        \%                              \\
Data/restraints/parameters                 & 243/0/8                              \\
Goodness-of-fit on \textit{F}$^{2}$        & 1.333                               \\
Final \textit{R} indices [\textit{I} ${>}$ 2sigma(\textit{I})]  & $R_{\rm 1}$ = 0.0270 \\ 
                                           & ${\omega}$$R_{\rm 2}$ = 0.1020 \\                    
\textit{R} indices (all data)              & $R_{\rm 1}$ = 0.0294 \\ 
                                           & ${\omega}$$R_{\rm 2}$ = 0.1039  \\
${\Delta}$$\rho_{\rm max}$ and ${\Delta}$$\rho_{\rm min}$ (e/$\AA^{3}$)    & 4.538 and -3.776 \\       
\hline         
                                                                        
  \end{tabular}
\end{table}
\begin{figure}[t!]
\includegraphics[width=0.7\linewidth]{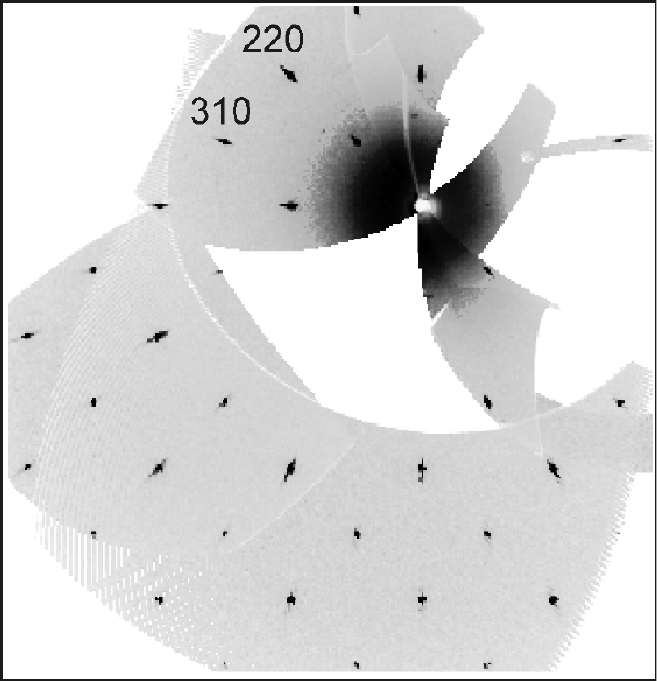}
\vspace{0cm}
\caption{ (Color online)  The reconstructed $\textit{hk0}$ reciprocal
space section of the single crystal EuFe$_{1.9}$Co$_{0.1}$As$_{2}$.} 
\label{fig7}
\end{figure}
structure was solved with XS \cite{XS} and subsequent structure
refinements were performed with XL. \cite{XL} Because of the almost equal
number of electrons, Co and Fe atoms were considered as one atom.
The final anisotropic full-matrix least-squares refinement on
${\textit F}$$_{0}$$^{2}$ with 8  variables converged at $R_{\rm 1}$ = 2.70 ${\%}$. 
Further details of the structure refinement are shown in Table~I.  No
additional phases (impurities, twins, or intergrowing crystals) were
detected by examination of the reconstructed reciprocal space
sections measured at room temperature (Fig.~2). In the
Co-substituted crystal the room-temperature lattice parameter  
${\textit c}$ is reduced and the lattice parameter ${\textit a}$ 
is increased relative to the parent compound
EuFe$_{2}$As$_{2}$ (${\textit a}$ = ${\textit b}$ = 3.898(1) \AA, ${\textit c}$ = 12.085(5) \AA) which was also grown out of Sn flux.\\

\subsection{Magnetic Properties}
\begin{figure}[b!]
\includegraphics[width=1\linewidth]{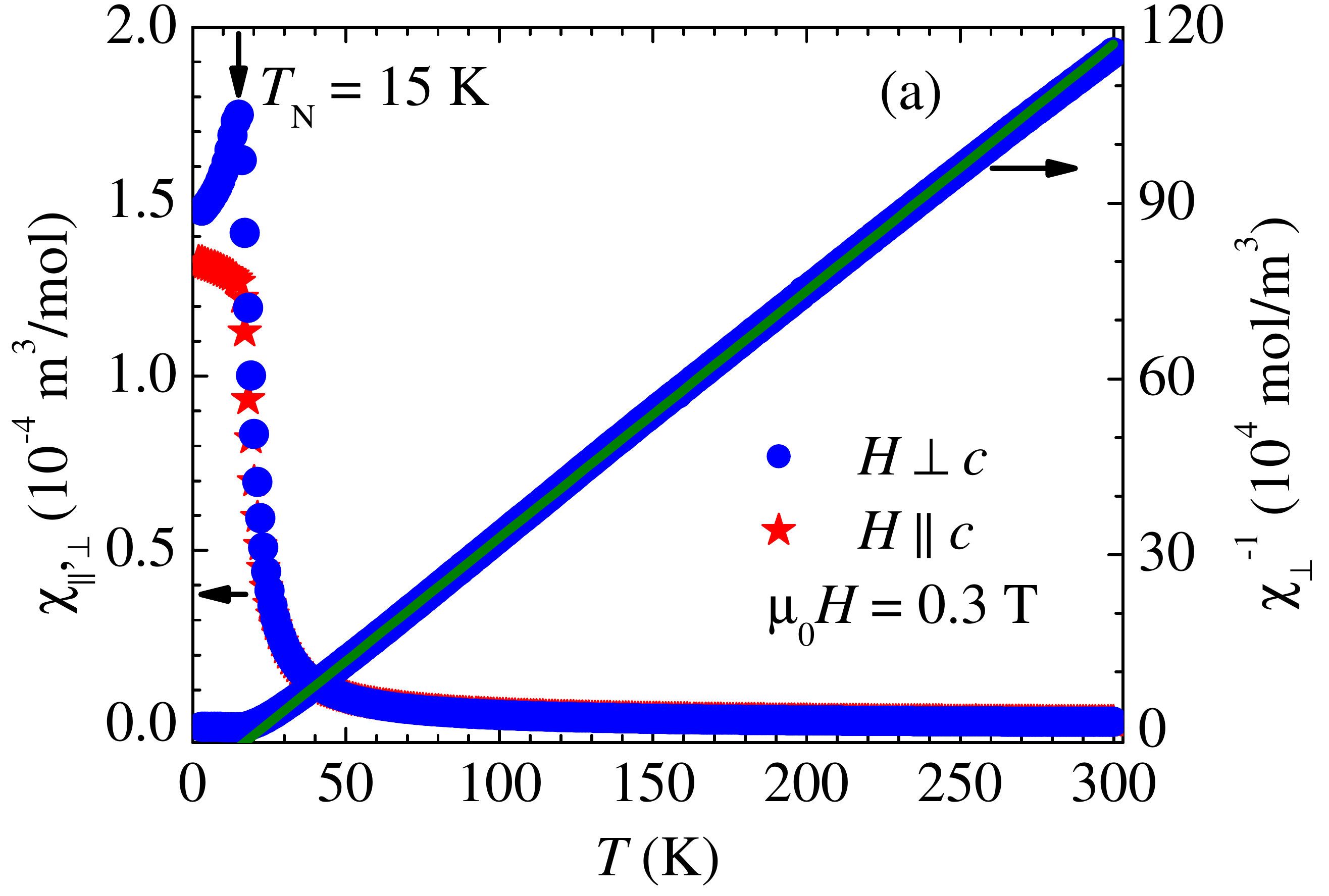}
\includegraphics[width=1\linewidth]{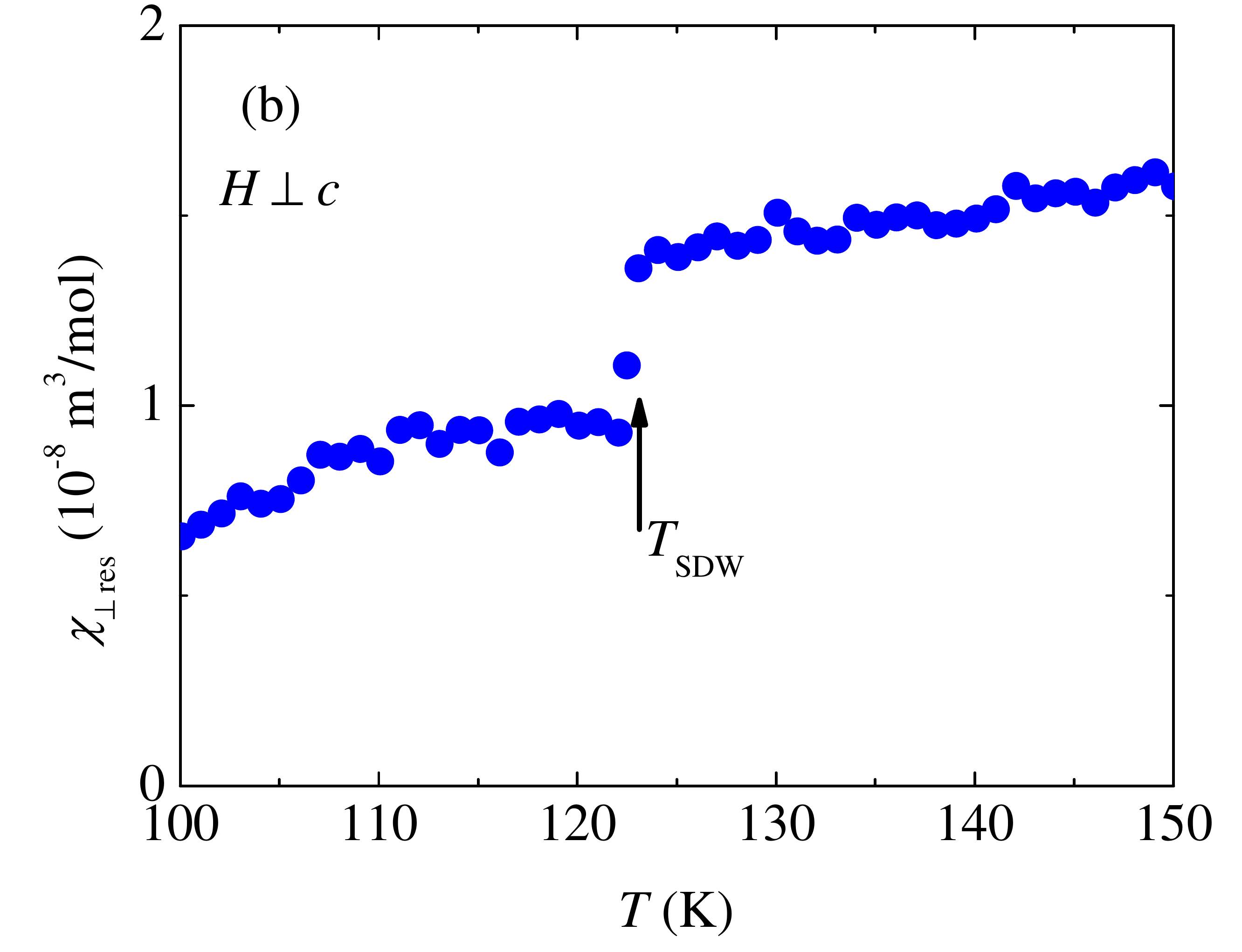}
\vspace{-0.7cm} \caption{ (Color online) (a) Temperature dependence of the magnetic
susceptibility measured in a field of ${\mu}$$_{0}$\textit{H} = 0.3 T applied parallel (\textit{H} ${\parallel}$ $\textit{c}$) and perpendicular (\textit{H} ${\perp}$ $\textit{c}$) to the crystallographic $c$-axis of the
single crystal EuFe$_{1.9}$Co$_{0.1}$As$_{2}$. The measurements
were performed in the zero-field cooling mode. In addition, for \textit{H} ${\perp}$ $\textit{c}$
the temperature dependence of the inverse susceptibility 1/${\chi}$$_{\perp}$ is plotted.
The solid line represents a fit to the data with the Curie-Weiss law given in Eq.(1). (b) Temperature dependence of   ${\chi}$$_{\perp}$$_{res}$(\textit{T})=${\chi}$$_{\perp}$(\textit{T})-${\chi}$$_{\perp}$$_{CW}$(\textit{T})   (see text for an explanation) in the single crystal EuFe$_{1.9}$Co$_{0.1}$As$_{2}$.} \label{fig2}
\end{figure}
  Figure~3a shows the temperature dependence of the magnetic susceptibility ${\chi}$  for the
EuFe$_{1.9}$Co$_{0.1}$As$_{2}$ crystal in an applied field of ${\mu}$$_{0}$\textit{H}=0.3 T 
parallel (\textit{H} ${\parallel}$ \textit{c}) and perpendicular (\textit{H} ${\perp}$ \textit{c}) to the crystallographic ${\textit c}$-axis. The temperature dependence of the inverse susceptibility for  \textit{H} ${\perp}$ \textit{c} is also shown. Below 30 K, ${\chi}$ increases sharply, indicating ferromagnetic coupling between the Eu$^{2+}$ moments. Below 15 K, a sudden decrease in the magnetic susceptibility for \textit{H} ${\perp}$ \textit{c} (${\chi}$$_{\perp}$) can be seen, indicative for a transition to the antiferromagnetic state of the Eu$^{2+}$ moments.  On the other hand, the magnetic susceptibility for \textit{H} ${\parallel}$ \textit{c} (${\chi}$$_{\parallel}$) remains almost constant below 15 K. This suggests that the Eu$^{2+}$ moments align in the ${\textit a}$${\textit b}$-plane, \cite{ShuaiJiang} similar as in the case of EuFe$_{2}$As$_{2}$. \cite{Xiao09} From 50 K to 120 K, ${\chi}$$_{\perp}$ is well described by the Curie-Weiss law: 
\begin{equation}
\chi_{\perp CW}(T)=\frac{C}{T+\Theta} \label{eq1}.
\end{equation}
Here ${\textit C}$ denotes the Curie-Weiss constant, and ${\Theta}$ the Weiss temperature. Analyzing the data with
Eq.(1) in the temperature range from 30 K to 120 K yields:
\textit{C}=2.43(5)${\times}$10$^{-4}$ m$^{3}$ K/mol and ${\Theta}$=-21.34(7) K. The Curie-Weiss
\textit{C} constant corresponds to an effective magnetic moment of
${\mu}$$_{eff}$ = 8.7 ${\mu}$$_{B}$, which is slightly larger than the theoretical value
of the magnetic moment of a free Eu$^{2+}$ ion (${\mu}$$_{free}$=7.94 ${\mu}$$_{B}$). 
The negative value of ${\Theta}$ infers that the interaction
between the Eu$^{2+}$ moments is ferromagnetic. Therefore, one can expect that the intralayer arrangment of the Eu$^{2+}$ spins is ferromagnetic as in the parent compound EuFe$_{2}$As$_{2}$. A clear drop in the susceptibility at the SDW transition temperature was observed in BaFe$_{2}$As$_{2}$. \cite{Rotter08}
In the case of EuFe$_{2}$As$_{2}$ the large signal from the Eu$^{2+}$
spins makes it impossible to directly observe the SDW
anomaly. However, after subtracting the Curie-Weiss 
contribution ${\chi}$$_{\perp}$$_{CW}$(\textit T) from ${\chi}$$_{\perp}$(\textit T), a small anomaly in 
${\chi}$$_{\perp}$$_{res}$(\textit T)=${\chi}$$_{\perp}$(\textit T) - ${\chi}$$_{\perp}$$_{CW}$(\textit T) is visible at around 120 K (Fig.~3b). This behavior resembles that observed in EuFe$_{2}$As$_{2}$ \cite{RenZhu} and BaFe$_{2}$As$_{2}$ \cite{Rotter08} which was ascribed to the SDW transition of the Fe moments.\\   

\subsection{Nuclear Magnetic Resonance}
 In this section we present $^{75}$As nuclear magnetic resonance (NMR) studies
in single crystal EuFe$_{1.9}$Co$_{0.1}$As$_{2}$. NMR is a powerful and
extremely sensitive microscopic tool to probe both, magnetism and the local structure in a solid.
$^{75}$As has a large quadrupolar moment (Q = 0.3b) that interacts
with the local electric field gradient (EFG) in the crystal. The
nuclear spin Hamiltonian describing the interactions of the
investigated nucleus with the external magnetic field and the crystal
electric field gradient at the nuclear site is given by the expression:
\begin{equation}
H = \gamma\hbar(1+K_\alpha)I_\alpha H_0+\frac{h\nu_Q}{6[(3I_z^2-1)+\eta(I_x^2-I_y^2)]}  \label{eq2}.
\end{equation}
 Here $K_{\rm \alpha}$ (${\alpha}$ =x, y, z) is the relative magnetic shift in the ${\alpha}$ direction,
$I_{\rm \alpha}$ (${\alpha}$ =x, y, z) are the nuclear spin components, $H_{\rm 0}$ is the external magnetic field, ${\gamma}$ is the gyromagnetic ratio, and $\nu_{\rm Q}$ is defined as:
\begin{equation}
\nu_Q = \frac{3eQV_{zz}}{2I(2I-1)h}
\end{equation}
 Where $V_{\rm zz}$  denotes the major principal axis of the EFG tensor, and ${\eta}$ the EFG
asymmetry parameter defined as ${\eta}$ = ($V_{\rm xx}$-$V_{\rm
yy}$)/{$V_{\rm zz}$} (0 ${\leq}$ ${\eta}$ ${\leq}$ 1). We use the
standard convention $V_{\rm xx}$ ${\leq}$ $V_{\rm
yy}$ ${\leq}$ $V_{\rm zz}$. Since the principal axis of the EFG
tensor as well as the magnetic shift tensor are defined by the
symmetry of the nuclear site, the resonance frequency of a
particular nuclear transition depends on the field
direction relative to the crystalline axes.

 In the absence of a static magnetic field, the remaining term gives
rise to double degenerate energy levels, between which nuclear
quadrupole resonance (NQR) transitions can be induced. $^{75}$As has
a nuclear spin \textit{I}=$\frac{3}{2}$ and thus two double degenerate
${\pm}$$\frac{1}{2}$ and ${\pm}$$\frac{3}{2}$  energy levels. In the
presence of a large external magnetic field $H_{\rm 0}$ a
splitting of the $^{75}$As spectrum into a central line,
arising from the central transition
(+$\frac{1}{2}$,-$\frac{1}{2}$) and two satellite lines due to
the (${\pm}$$\frac{1}{2}$,${\pm}$$\frac{3}{2}$) transitions occurs.
A representative $^{75}$As NMR spectrum of the central
transition at 295 K
is shown in the inset of Fig.~4. In the paramagnetic state (${\textit
T}$ ${\\>}$ $T_{\rm SDW}$) lines with a full width at half
maximum (FWHH) of about 250 kHz (central line) and 500 kHz (satellite lines) are observed. 
\begin{figure}[t!]
\includegraphics[width=1\linewidth]{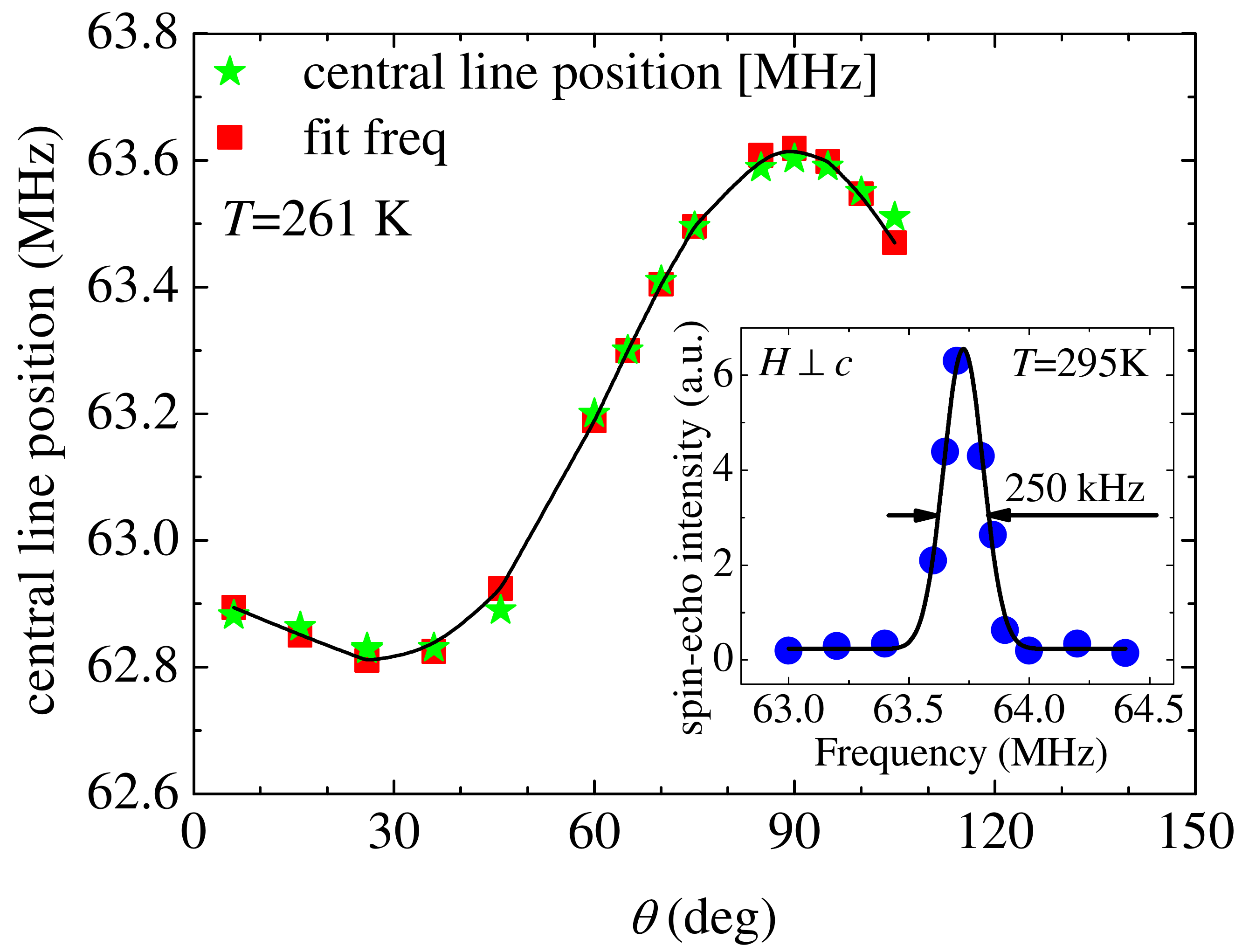}
\caption{ (Color online) Dependence of the $^{75}$As NMR central line
frequency at 261 K on the angle ${\theta}$ between the external magnetic field
and the crystallographic \textit{c}-axis of single crystal EuFe$_{1.9}$Co$_{0.1}$As$_{2}$.
The solid line is a guide to the eye. 
Solid squares represent the calculated frequencies as described in the text. The inset illustrates the $^{75}$As NMR central line shape at 295 K. The solid line
represents a Gaussian fit.} 
\label{fig3}
\end{figure}
Figure~4 presents the dependence of the $^{75}$As-NMR central line position
on the angle ${\theta}$ between the magnetic field orientation and the
\textit{c}-axis at 261 K (tetragonal phase). For all orientations of the magnetic field with
respect to the \textit{c}-axis the line positions show a strong
negative shift relative to the corresponding $^{75}$As Larmor
frequency of 65.9 MHz at 9 T.
$^{75}$As has four nearest neighbor Fe atoms, and lies just above
or below the Fe plane (see Fig.~1). In the tetragonal phase its site
symmetry requires uniaxial symmetry along the \textit{c} axis for the EFG
tensor(${\eta}$ = 0) and the $K$ tensor. The analysis of the angular dependence of the frequency of the central line using the diagonalization of the Hamiltonian [Eq.~(2)] yields
$K_{\rm ab}$ = -0.0372(2), $K_{\rm c}$ = -0.0456(3), and $\nu_{\rm
Q}$ = 7.39(24) MHz with ${\eta}$ set to be zero.

 Figure~5a shows the temperature dependence of the spin-spin relaxation rate 1/$T_{\rm 2}$
of the central line of the $^{75}$As NMR spectrum for \textit{H} ${\parallel}$ \textit{c}.
\begin{figure}[t!]
\includegraphics[width=1\linewidth]{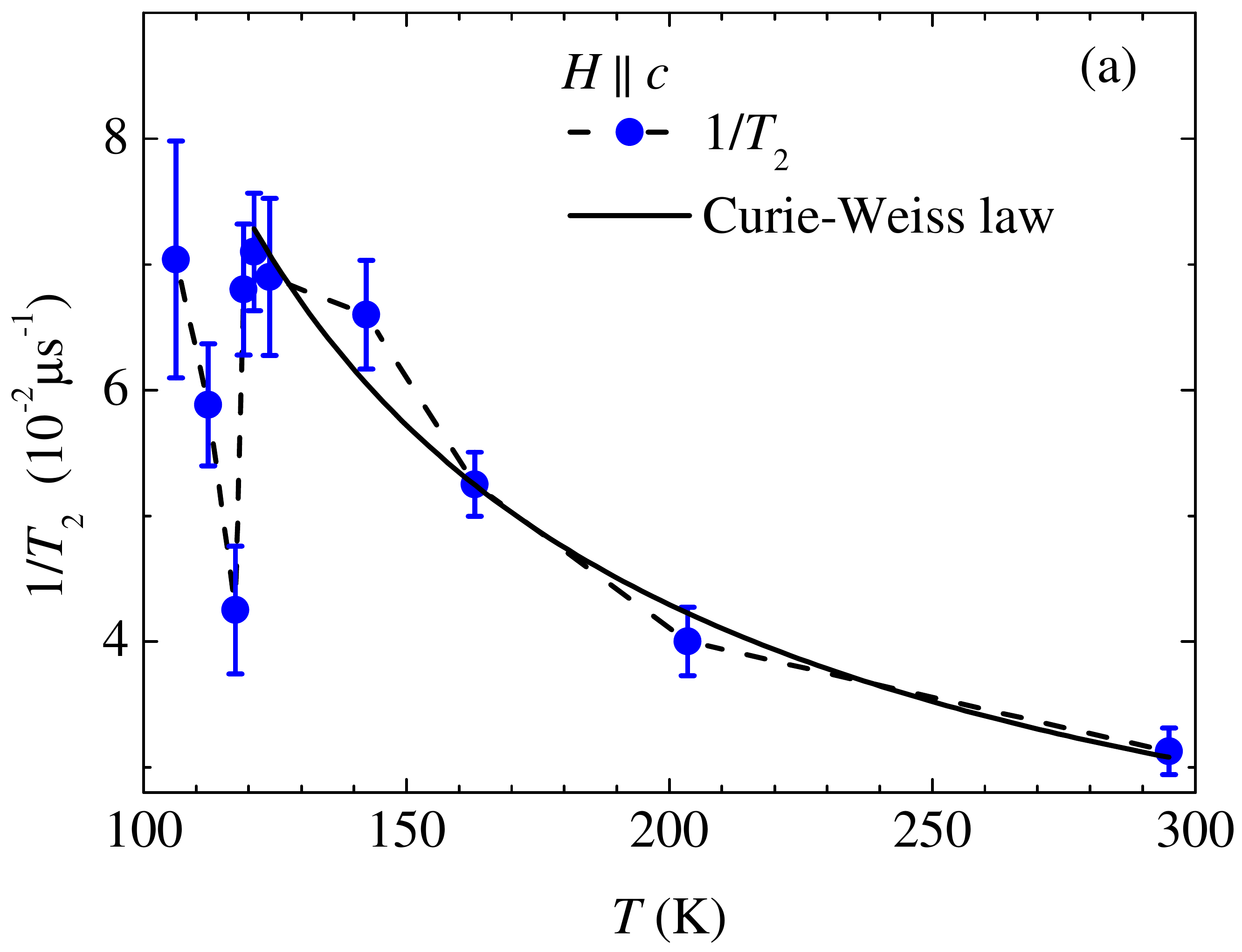}
\includegraphics[width=1\linewidth]{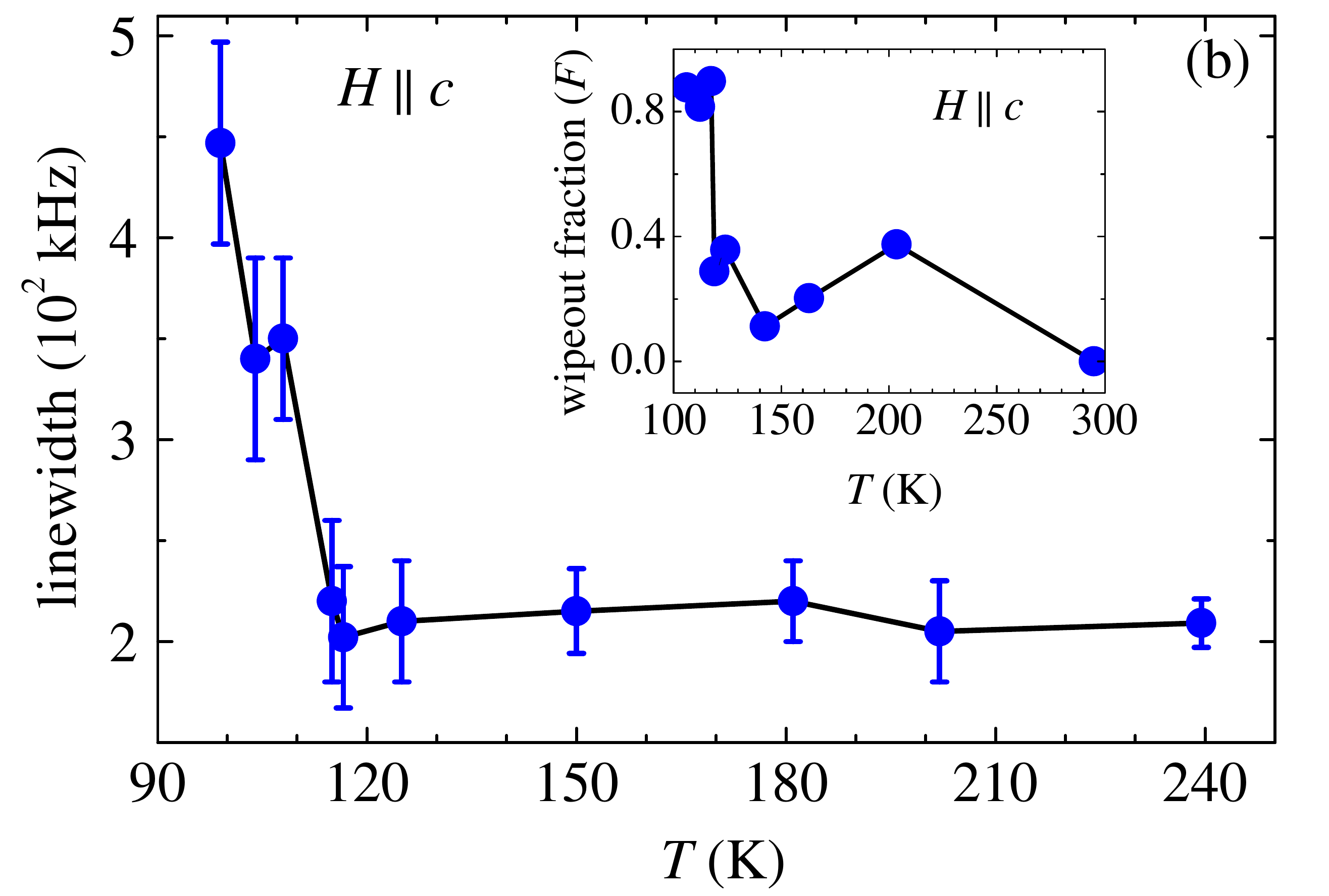}
\vspace{0cm} \caption{ (Color online) 
(a) Temperature
dependence of the spin-spin relaxation rate 1/$T_{\rm 2}$ of the
$^{75}$As central line of the single crystal EuFe$_{1.9}$Co$_{0.1}$As$_{2}$ for \textit{H} ${\parallel}$ $\textit{c}$ . The solid line represents the Curie-Weiss law, and the dashed line is to guide to the eye.  (b)
Temperature dependence of the linewidth (FWHM) and the wipeout fraction $F$ 
(inset) for \textit{H} ${\parallel}$ $\textit{c}$. Solid lines are guides to the eye.} 
\label{fig5}
\end{figure}
1/$T_{\rm 2}$ exhibits a Curie-Weiss-like
temperature dependence down to 120 K. This
shows that the $^{75}$As nuclei interact with the localized Eu-4$f$
moments. A
reduction of 1/$T_{\rm 2}$ is observed at 120 K, which reflects the slowing down of the Fe spin
fluctuations due to the SDW ordering of the Fe moments. A similar behavior across the SDW transition
was also reported for spin-lattice
relaxation measurements in \textit{A}Fe$_{2}$As$_{2}$ (\textit{A}=Ba, Ca, Sr). \cite{Kitagawa1,Kitagawa2,Ning08}  However, in contrast to these findings, in our case 1/$T_{\rm 2}$ increases again upon further
cooling. This increase reflects the dominant Eu$^{2+}$ contribution in the
spin-spin relaxation process. The SDW transition below 120 K is also reflected
in the temperature dependence of the NMR linewidth as shown in Fig.~5b.
In addition, we observed a
so-called wipeout phenomenon of the central line intensity with
decreasing the temperature across the SDW transition.
The temperature dependence of the wipeout fraction, defined as
${\textit{F}}$ = [$A_{\rm 1}$(295 K) - $A_{\rm 1}$(T)]/$A_{\rm
1}$(295 K) is displayed in the inset of Fig.~5b.
The wipeout fraction is a measure for an unobserved signal
intensity, \cite{Bloembergen53,Winter71}  $A_{\rm 1}$=\textit{I} ${\cdot}$ T, where \textit{I} represents the NMR echo intensity, corrected for the $T_{\rm 2}$ echo decay and integrated over the full central
line. A sudden increase in the wipeout fraction and a pronounced
broadening (Fig.~5b) of the $^{75}$As central line can
be clearly seen below 120 K.
The decrease of 1/$T_{\rm 2}$, the pronounced broadening of the $^{75}$As
central line, and the sudden jump of the wipeout fraction are caused by the 
appearance of inhomogeneous internal magnetic fields in the SDW phase.
 
 Next we describe the determination of the hyperfine coupling
strength between the $^{75}$As nuclei and Eu$^{2+}$ 4$f$ moments. For this reason we measured the temperature dependence
of the magnetic shift \textit{K} of the central line of the $^{75}$As NMR spectrum.
Since for this temperature range $\nu_{\rm Q}$ is
almost constant, as it will be shown below and ${\eta}$ is very close to zero, the observed
temperature dependence of the central line frequency is fully
determined by the temperature behavior of \textit{K}. In Fig.~6a we
present the temperature dependence of the shift \textit{K} in the
temperature range from 100 and 300 K for ${\textit{H}}$ ${\parallel}$ $c$ ($K_{\rm \parallel}$=$K_{\rm c}$) and
${\textit{H}}$ ${\perp}$ $c$ ($K_{\rm \perp}$=$K_{\rm ab}$). Compared to the
magnetic shift data for \textit{A}Fe$_{2}$As$_{2}$ (\textit{A}=Ba, Ca, Sr), \cite{Kitagawa1,Kitagawa2,Ning08,Baek09} our
\begin{figure}[b!]
\includegraphics[width=1\linewidth]{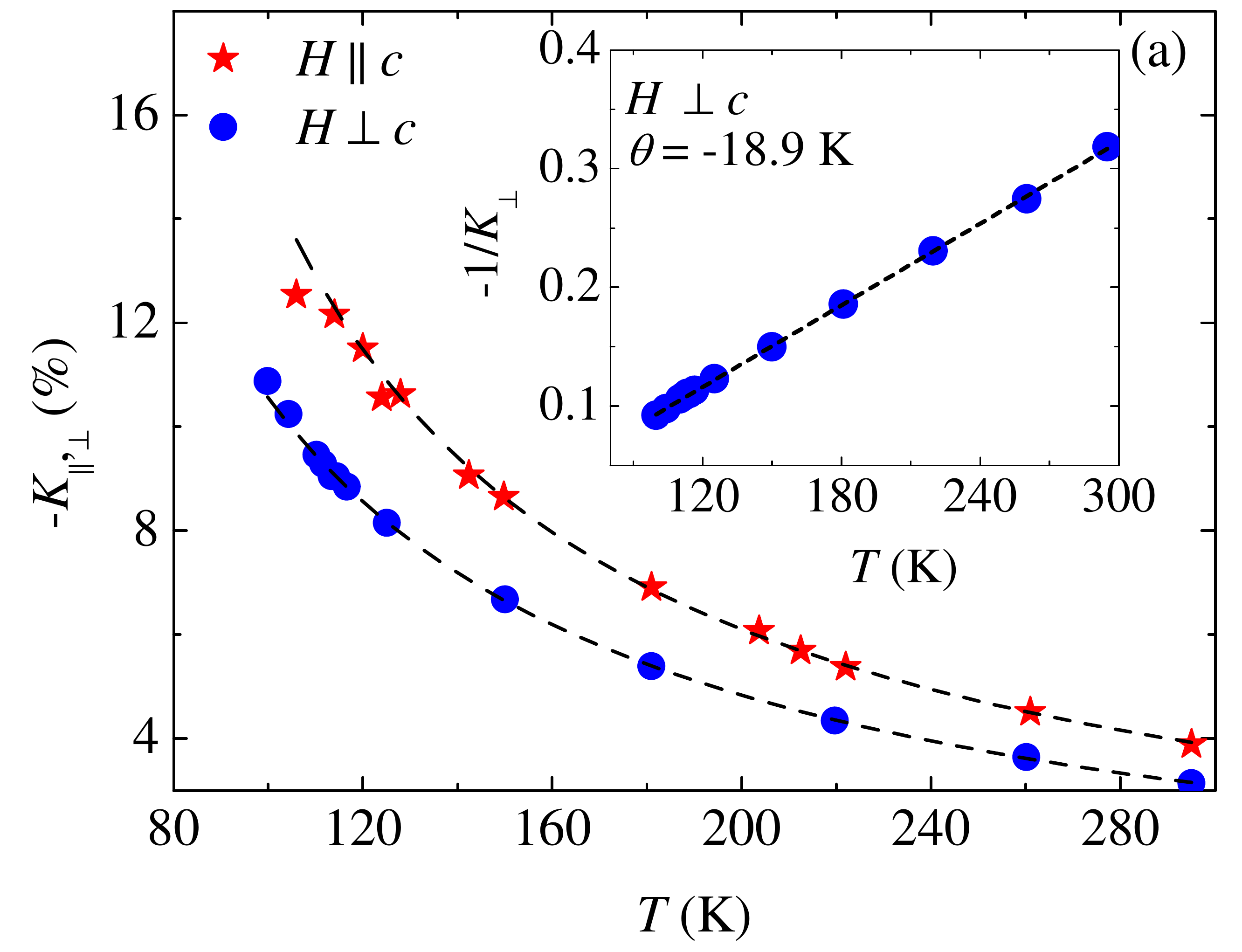}
\vspace{0cm}
\includegraphics[width=1\linewidth]{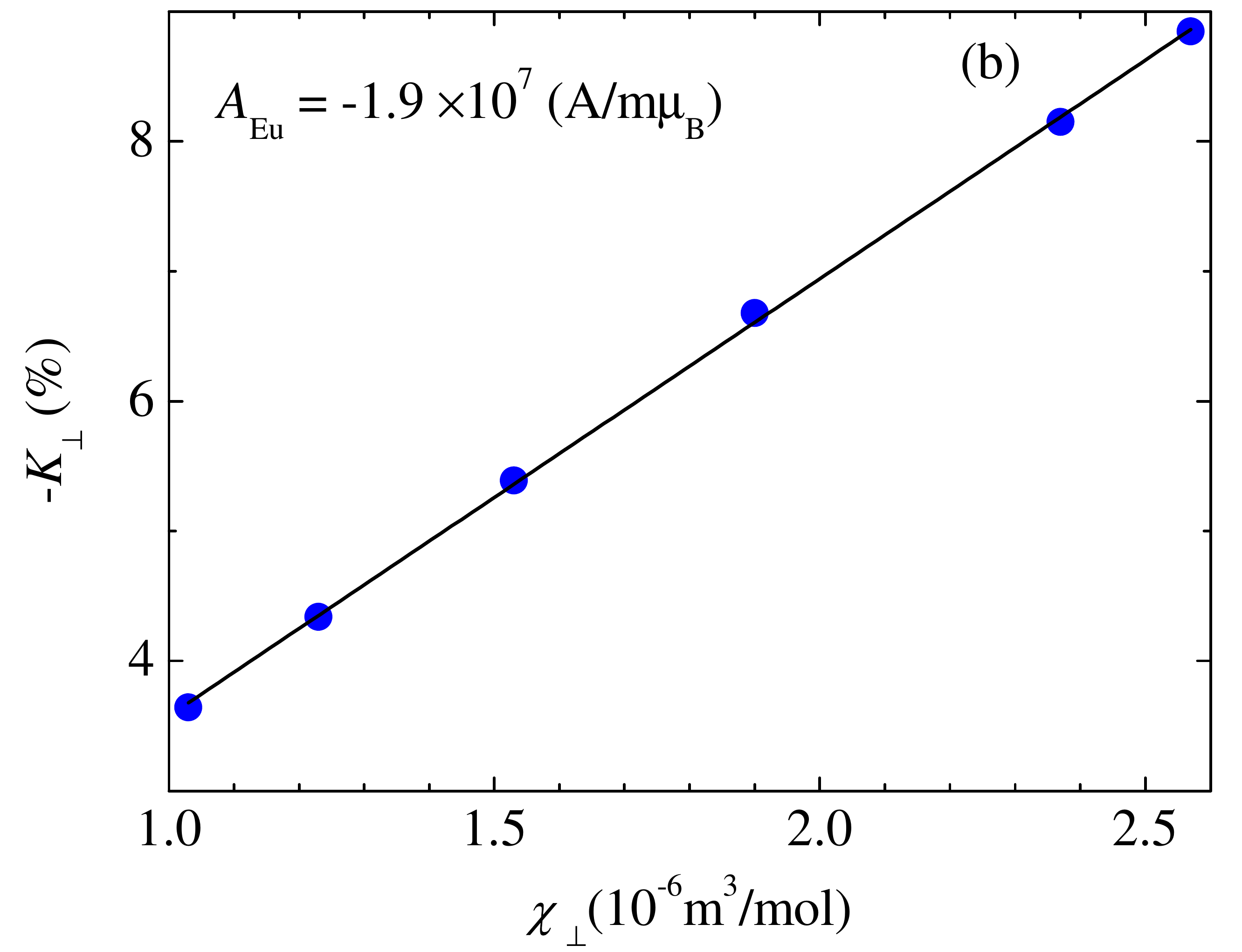}
\vspace{-0.5cm} \caption{ (Color online) (a) Temperature dependence of the
$^{75}$As magnetic shift in a single crystal EuFe$_{1.9}$Co$_{0.1}$As$_{2}$ for \textit{H} ${\parallel}$ $\textit{c}$ and \textit{H} ${\perp}$ $\textit{c}$. The inset shows the inverse of the temperature dependent part of the shift 
-1/$K_{\rm \perp}$ as a function of \textit{T}. The dashed lines represent the Curie-Weiss behavior. (b) Plot of $-K_{\rm \perp}$ vs. $\chi_{\rm \perp}$
as obtained from the susceptibility measurements. The solid line is a linear fit.}
\label{fig4}
\end{figure}
observed shifts are significantly larger, negative,
and show a completely different temperature dependence.
The temperature dependence of the relative magnetic shift \textit{K} above 117 K 
is well described by a Curie-Weiss like behavior for both directions of the magnetic
field \textit{H}:  
\begin{equation}
K(T) = K_{0}+\frac{C_K}{T + \Theta} \label{eq3}.
\end{equation}
 The inset of Fig.~6a presents the inverse of
the temperature dependent part of the shift for ${\textit{H}}$ ${\perp}$ $c$ as a function of temperature.
Below the SDW transition $T_{\rm SDW}$ = 120 K the data deviates from the Curie-Weiss behavior. This deviation can be understood when considering that above $T_{\rm SDW}$ both the Eu and Fe sublattices are in the paramagnetic state,
and both contribute to the shift, while below $T_{\rm SDW}$
only the Eu$^{2+}$ moments contribute.  An analysis of the
data for \textit{H} ${\perp}$ \textit{c} using Eq.~(4) yields:
\textit{${\Theta}$} = -18.9(9) K, $K_{\rm 0}$ = 0.17(25) ${\%}$.
The value of ${\Theta}$ is in fair agreement with the value ${\Theta}$ = -21.34(7) K
determined from the magnetic susceptibility measurements. This suggests
that the Curie-Weiss part of the shift arises from the
hyperfine coupling between the $^{75}$As nuclei and the Eu$^{2+}$ 4$f$ moments.
The remaining constant part of the shift $K_{\rm 0}$ could be
related to the coupling of $^{75}$As with the itinerant 3d electrons
in the Fe$_{2}$As$_{2}$ layer, including an orbital shift. However, the value of
$K_{\rm 0}$ is a factor two smaller than the total $^{75}$As
magnetic shift reported for BaFe$_{1.8}$Co$_{0.2}$As$_{2}$. \cite{Baek09}
The contribution of Eu to the magnetic shift $K_{\rm Eu}$ can be
related to the susceptibility ${\chi}$$_{Eu}$  of the localized Eu 4$f$ moments
as follows:
\begin{equation}
\ K_{\rm Eu} = \frac{A_{Eu}}{gN_{A}\mu_{B}}\chi_{\rm Eu} \label{eq2}
\end{equation}
 where $A_{\rm Eu}$ is the $^{75}$As hyperfine coupling with the 4$f$ moments, $N_{\rm A}$ and ${\mu}$$_{\rm B}$ are the
Avogadro number and Bohr magneton, respectively. Fig.~6b shows $K_{\rm \perp}$ versus
${\chi}_{\rm \perp}$ with the temperature as an implicit parameter.
From the linear fit of this data we can estimate a hyperfine
coupling constant $A_{\rm Eu}$ = -1.9${\times}$10$^{7}$ A/m per $\mu_{\rm B}$. This value of \textit{A} is
almost 60 times larger than the one reported for
NdFeAsO$_{0.85}$F$_{0.15}$. \cite{Jeglie} It is known that the '1111' compounds are
more anisotropic than the '122' compounds. \cite{Fengjie10} Therefore, the
'1111' systems are treated as quasi two-dimensional, while the
'122' systems are regarded as three dimensional systems. Moreover, the
distance between the rare-earth ion Eu$^{2+}$ and the conduction
layers in EuFe$_{1.9}$Co$_{0.1}$As$_{2}$ is \textit{d} = \textit{c}/2 = 5.9713(3)
\AA, which is much smaller than in
NdFeAsO$_{0.85}$F$_{0.15}$ (\textit{d} = 8.577 \AA). \cite{Jeglie} The smaller interlayer distance
in the '122' compounds as compared to the '1111' system and the more isotropic band structure may be the reason for the much stronger hyperfine coupling in the '122' compound. The large value of hyperfine coupling
constant, quantitatively determined from the present NMR experiment, provides direct experimental evidence for a
strong coupling between the Eu$^{2+}$ localized moments and the Fe$_{1.9}$Co$_{0.1}$As$_{2}$ layers. 
This suggests that the magnetic
exchange interaction between the localized Eu 4$f$ moments is mediated by the itinerant Fe 3$d$ electrons, \textit{i.e.}, via a Ruderman-Kittel-Kasuya-Yosida (RKKY) type of mechanism \cite{Raffius93,ZRen09} leading to a high magnetic ordering temperature of the Eu$^{2+}$ moments in EuFe$_{2}$As$_{2}$. It was shown that  Co-substitution induces superconductivity in EuFe$_{2-x}$Co$_{x}$As$_{2}$ with a
reentrant behavior of resistivity due to the antiferromagnetic ordering of the
Eu$^{2+}$ spins. \cite{He08}
Reentrant superconducting behavior was also observed in 
a EuFe$_{2}$As$_{2}$ crystal under an applied pressure up to 2.5 GPa. \cite{Miclea09,Terashima} Moreover, while
superconductivity with a maximum $T_{\rm c}$ = 20 K has
been reported for BaFe$_{2-x}$Ni$_{x}$As$_{2}$, \cite{LiLuo08} no superconductivity was observed in EuFe$_{2-x}$Ni$_{x}$As$_{2}$. \cite{ZRen09} The
strong interaction between the localized Eu$^{2+}$ moments and charge carriers in the Fe$_{2-x}$Co$_{x}$As$_{2}$ layers  
may cause pair breaking according to the Abrikosov-Gorkov theory, \cite{Abrikosov} which may be the reason why it is difficult to get superconductivity in EuFe$_{2-x}$Co$_{x}$As$_{2}$.

 Finally, we would like to discuss the temperature dependence of the resonance frequency
and the skewness of a $^{75}$As lower-satellite line for
\textit{H} ${\perp}$ \textit{c} (see Fig.~7). 
\begin{figure}[b!]
\includegraphics[width=1.55\linewidth]{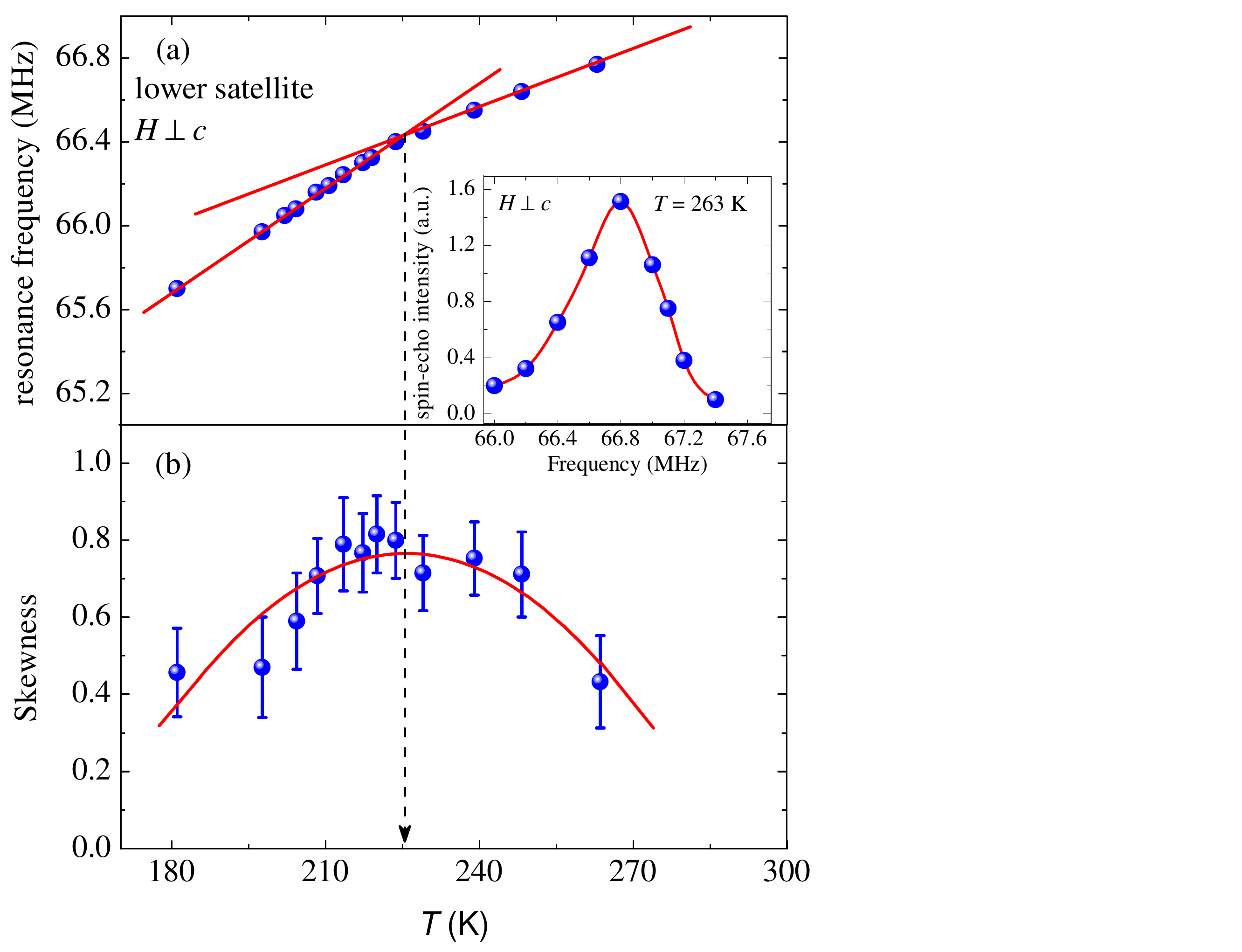}
\caption{(Color online)
Temperature dependence of the frequency (a) and the skewness (b) of the $^{75}$As lower-satellite line
for \textit{H} ${\perp}$ \textit{c}. The inset illustrates 
the $^{75}$As NMR lower-satellite line shape at 263 K for \textit{H} ${\perp}$ \textit{c}.
Solid and dashed lines are guides to the eye.}
\label{fig3}
\end{figure}
A Change in the slope of the temperature 
dependence of the satellite line frequency can be seen at 225 K
as shown in Fig.~7a.
The inset of Fig.~7a illustrates a
typical $^{75}$As-NMR spectrum of the low-frequency satellite at 263 K.
The spectra exhibit an asymmetric lineshape in the 
investigated temperature range. 
We analyzed the asymmetry of the satellite line shape at different temperatures
by calculating its skewness (see Fig.~7b),
defined as the third standartized moment. With decreasing temperature
the skewness exhibits an increase down to 225 K, where it starts to reduce again. 
The slope change in the temperature dependence of the frequency of the $^{75}$As lower-satellite line as well as the maximum in the temperature evolution of the skewness may be related to the appearence of a electronic nematic phase below 225 K in EuFe$_{1.9}$Co$_{0.1}$As$_{2}$. Note that evidence for an electron nematic phase transition was recently established in undoped \textit{A}Fe$_{2}$As$_{2}$ (\textit{A}=Ba, Ca) \cite{Tanatarv3} and Co-doped
BaFe$_{2-x}$Co$_{x}$As$_{2}$ \cite{ChuAnalytis} single crystals by means of in-plane anisotropy
measurements of the electrical resistivity.

 In order to get more quantitative NMR results for a possible nematic
phase, the angular dependence of the full $^{75}$As NMR spectrum
(central line and both satellite lines) was measured at 181 K. 
The dominant contribution to the magnetic shift stems from the Eu$^{2+}$
4$f$ moments (see above), and it is acceptable to assume that this
contribution does not have a strong
\textit{a}\textit{b}-anisotropy.
Therefore, we reduced the parameter set to $K_{\rm a}$ = $K_{\rm b}$ = $K_{\rm ab}$, $K_{\rm
c}$, $\nu_{\rm Q}$, and ${\eta}$. Analysis of the angular dependence of the full spectrum using the diagonalization  of the Hamiltonian [Eq.~(2)] yields $K_{\rm ab}$ =
-0.059(1), $K_{\rm c}$ = 0.069(2), $\nu_{\rm Q}$ = 7.51(17) MHz, and
${\eta}$ = 0.04(3). The change in $\nu_{\rm Q}$ compared to
the value $\nu_{\rm Q}$ = 7.39(24) MHz obtained at 261 K in the tetragonal phase is small.
The slight increase of ${\eta}$ may reflect the lower symmetry of the As site in a possible nematic phase. However, further experiments are needed to clarify the presence of a nematic phase in EuFe$_{1.9}$Co$_{0.1}$As$_{2}$. Resistivity measurements on detwinned\cite{Tanatar09} single crystals could provide more information by probing the in-plane electronic anisotropy.

\subsection{SUMMARY AND CONCLUSIONS}
In summary, the 
magnetic properties of 
EuFe$_{1.9}$Co$_{0.1}$As$_{2}$ single crystal were investigated by X-ray
diffraction, magnetization, and $^{75}$As NMR experiments. It was found
that the temperature dependence of the $^{75}$As magnetic
shift as well as the spin-spin relaxation rate follow a Curie-Weiss
type behavior, implying that the $^{75}$As nuclei interact with the
localized Eu 4$f$ moments in the Eu layer. A large value of the
hyperfine coupling constant between the $^{75}$As nuclei and the Eu 4$f$ moments suggests a strong coupling between the Eu and
Fe$_{1.9}$Co$_{0.1}$As$_{2}$ layers. Due to such a strong interlayer coupling
the antiferromagnetic interaction between the localized Eu$^{2+}$ 4$f$ moments is probably mediated by
a Ruderman-Kittel-Kasuya-Yosida (RKKY) type interaction. Evidence for a
SDW transition at 120 K was obtained from magnetic susceptibility as
well as from $^{75}$As-NMR measurements. A change
of the slope in the temperature dependence
of the frequency of the $^{75}$As lower-satellite line is observed at 225 K. In addition, at the same temperature
also a maximum in the temperature behavior of the skewness is detected. These findings may
indicate a phase transition to an electron nematic state below 225 K.\\ 

\section{Acknowledgments}~
~This work was supported by the Swiss National Science Foundation, the
SCOPES grant No. IZ73Z0${\_}$128242, the NCCR Project MaNEP, the EU Project
CoMePhS, and the Georgian National Science Foundation grant
GNSF/ST08/4-416.

\end{document}